\documentclass[journal]{IEEEtran}
\usepackage{amsmath}
\usepackage{graphicx}
\usepackage{multirow}
\usepackage{float}
\usepackage{comment}
\ifCLASSINFOpdf
\else
\fi
\begin{document}
%
\title{Reinforcement Learning for Traffic Control with Adaptive Horizon}
%
%
%

\author{Wentao~Chen,
        Tehuan~Chen,
        and~Guang~Lin
\thanks{Wentao Chen is with the School of Mechanical Engineering, Purdue University, West Lafayette, Indiana 47907, USA. email: chen1853@purdue.edu}
\thanks{Tehuan Chen is with the School of Mechanical Engineering and Mechanics, Ningbo University, Ningbo, Zhejiang 315211, China}
\thanks{Guang Lin is with the Department of Mathematics and School of Mechanical Engineering and Department of Statistics (by courtesy), Purdue University, West Lafayette, Indiana 47907, USA.}}

%
%


\markboth{}%
{Chen \MakeLowercase{\textit{et al.}}: Reinforcement Learning for Traffic Control with Adaptive Horizon}
%



\maketitle

\begin{abstract}
This paper proposes a reinforcement learning approach for traffic control with the adaptive horizon. To build the controller for the traffic network, a Q-learning-based strategy that controls the green light passing time at the network intersections is applied. The controller includes two components: the regular Q-learning controller that controls the traffic light signal, and the adaptive controller that continuously optimizes the action space for the Q-learning algorithm in order to improve the efficiency of the Q-learning algorithm. The regular Q-learning controller uses the control cost function as a reward function to determine the action to choose. The adaptive controller examines the control cost and updates the action space of the controller by determining the subset of actions that are most likely to obtain optimal results and shrinking the action space to that subset. Uncertainties in traffic influx and turning rate are introduced to test the robustness of the controller under a stochastic environment. Compared with those with model predictive control (MPC), the results show that the proposed Q-learning-based controller outperforms the MPC method by reaching a stable solution in a shorter period and achieves lower control costs. The proposed Q-learning-based controller is also robust under 30\% traffic demand uncertainty and 15\% turning rate uncertainty. 
\end{abstract}


%
\IEEEpeerreviewmaketitle

\section{Introduction}
%
%
%
%


Traffic congestion often occurs due to increased traffic volume. The normal operation of the whole society and economy are greatly affected. For instance, in China, the economic losses caused by traffic congestion in 15 large and medium-sized cities, including Beijing and Shanghai, are close to 1 billion each day in 2012. Thus, efficient vehicle operation and management is critical. Advanced Traffic Signal Control System (ATSCS),\cite{gartner1995development,singh2009time}, an efficient and convenient control system with artificial intelligence, is given a lot of attention from researchers. 

Generally, there are three mainstream control strategies in ATSCS: isolated intersection control\cite{yang2016isolated}, fixed time coordinated control\cite{papageorgiou2003review} and coordinated traffic-responsive control\cite{ye2016hierarchical}. An improved adaptive control method consists of a vehicle arrival estimation model and a signal optimization algorithm.\cite{chen2016improved} In this method, a schedule-driven intersection control strategy is built with the structural information in non-uniformly distributed traffic flow.\cite{xie2012schedule} However, since the adjacent intersections are neglected by the isolated intersection control, the application of isolated intersection control is greatly limited. Considering the drawback of the isolated intersection control, researchers developed an alternative fixed-time coordinated control strategy. In the fixed- time coordinated control strategy, researchers used a new two-direction green wave intelligent control strategy to solve the coordination control problem of urban arterial traffic.\cite{kong2011urban} Lu et al. also applied the coordination methodology for arterial traffic signal control based on a novel two-way bandwidth maximization model.\cite{lu2011two} However, when facing real-time traffic variations, fixed-time coordinated control strategies is invalid. Due to the advantage of real-time and flexibility, researchers proposed a coordinated traffic-responsive control strategy. Wang et al. adopted parallel control and management for intelligent transportation systems.\cite{wang2010parallel} Aboudolas investigated the rolling-horizon quadratic-programming approach for real-time network-wide signal control in large-scale urban traffic networks.\cite{aboudolas2010rolling}.  This paper considers the reinforcement learning for traffic signal control based on the coordinated traffic-responsive control.
 
Reinforcement learning (RL) is a machine learning algorithm in which the agent chooses the appropriate action to maximize the reward in a particular environment. \cite{kaelbling1996reinforcement}. In recent years, RL has been applied to traffic management by many researchers, in order to improve the performance of coordinated urban traffic control. El-tantawy et al. has reviewed the application of reinforcement learning on traffic light control from 1997 to 2010, which includes Q-learning and SARSA.\cite{el2014design} Li et al. reduce traffic congestion at freeway bottleneck by applying Q-learning-based speed limit.\cite{li2017reinforcement} In the past two years, researchers start to apply deep reinforcement learning to traffic control. Aslani et al. proposed building the Reinforcement Learning-embedded traffic signal controllers(RLTSCs) for traffic systems with sensor noise and disturbance, and tested the system with Q-learning, SARSA and actor-critic methods.\cite{aslani2018traffic} In the traffic signal control strategy proposed by Mannion, parallel computing is combined with Q-learning to increase the exploration efficiency and decrease delay time and queue length.\cite{mannion2015parallel} In recent two years, deep reinforcement learning has been proposed that combines deep learning and reinforcement learning to save the memories for storing reward function values. Wei applied deep Q-learning (DQN) to real-world data obtained from surveillance cameras.\cite{wei2018intellilight} Huang et al. proposed multimedia traffic control with deep reinforcement learning.\cite{huang2018deep} However, the previous works in reinforcement learning-based traffic control fail to take the change of action space into consideration. As the training goes on, we want the learning agent to be able to recognize a subset of action space as the most effective action set. With an adaptive action space, the agent does not have to explore adopt the actions that are deemed as ineffective and will save a significant amount of computational resources, which can make the controller more responsive to the traffic conditions. 

The rest of this paper is organized as follows. In Section 2, action, state and rewards for the traffic control are defined and the control problem for the traffic control is investigated. In Section 3, the Q-learning algorithm combined with adaptive horizon for traffic control is proposed. Then, the traffic control with the uncertainty of the traffic demand and turning rate is discussed. In Section 4, numerical simulation carries out to verify the advantage of the proposed control strategy compared to convention model predict control from the perspective of the cost function. Moreover, the proposed strategy shows robustness when facing uncertainty for the traffic demand and turning rate within a certain range of values.


\section{Q-learning for Traffic Control}
 In this research the RL agent interacts with the environment in discrete time steps, following the Markovian Decision Process (MDP) $\mathcal{M} = <\mathcal{S},\mathcal{A},\mathcal{P},\mathcal{R},\gamma>$, where $S$ is the state space that contains all states, $A$ is the action space with all possible actions, and $R$ is the reward for each state-action combination at each time step, $\mathcal{P}$ is the state transition probability function and $\gamma$ is the discount factor. In the interaction with the MDP, the agent generates a sequence of states, action and rewards. $H = <S_{1},A_{1},R_{1},S_{2},A_{2},R_{2}...>, S_{t} \in S, A_{t} \in A, R_{t} \in \mathbf{R}$,  $\mathbf{P} [S_{t+1} = s'|S_{t} = s,A_{t} = a] = \mathcal{P}(s,a,s')$ and $\mathbf{E} [R_{t}|S_{t} = s, A_{t} = a] = \mathcal{R}(s,a)$. 

Assuming discounted reward over time, the reward at each time step $R_{t}$ is based on the reward at previous time step $R_{t-1}$ and the discount factor $\gamma$ (0 $\leq$ $\gamma$ $\leq$ 1). The cumulative reward is represented as
\begin{equation}
 G = \sum_{t = 0}^{\infty}  \gamma^{t}R_{t}     
\end{equation}

To avoid traffic congestion, the optimal green light duration at each intersection should be chosen. The action space of the agent is to adopt different combinations of green light duration at each intersection. Each time an action is taken, the state changes. Given the current state s, the agent determines which action to take according to a policy $\pi_{\phi}(a|s)$ to maximize $G$, which is the probability that the agent choose action given state s. With $\mathcal{P}$ and $\mathcal{P}_{\theta}$ being parameters and $O(H)$ being the MDP objective, the agent achieves its own objective

\begin{equation}
\begin{split}
    \theta^{*} = \operatorname*{arg\,max}_\theta \mathbf{E}[& O(H)|\mathcal{M}_{\theta} = <\mathcal{S},\mathcal{A},\mathcal{P},\mathcal{R},\gamma>, \\
    & \pi_{\phi^{*}} = \operatorname*{arg\,max}_{\pi_{\phi}} \mathbf{E}[G|\pi_{\phi},\mathcal{M}_{\theta}]].
\end{split}
\end{equation}

\subsection{Q-learning algorithm}
The Q-learning algorithm is one of the fundamental RL algorithms. \cite{watkins1992q} The state set, action space, and reward function are determined for the Q-learning agent. At each discrete time step, the agent observes the current state and takes action to another state that maximizes the reward. To obtain the optimal result, the agent follows a value function mapping from the state space to the action space, which is a policy that guides the agent's behavior. By looking at the reward of each action, the agent determines which action should be chosen to obtain the highest cumulative rewards over time. A Q-table that contains the Q-values of all state-action combinations is established. The definition of Q-value can be represented as:
\begin{equation}
Q:S \times A \to R   
\end{equation}
 
The Q-value at each time step is updated with the rule 
\begin{equation}
\begin{split}
 Q^{t+1}&(s_{t},a_{t}) = Q^{t}(s_{t},a_{t})+ \\
                      & \epsilon[R_{t+1}+\gamma\cdot (max Q^{t}(s_{t+1},a_{t+1})- Q^{t}(s_{t}.a_{t}))]   
\end{split}
\end{equation}
where $Q^{t}(s_{t},a_{t})$ is the Q-value at time t for state $s_{t}$ and action $a_{t}$. $\epsilon$ is the learning rate that describes the likelihood of exploitation in the $\epsilon$-greedy algorithm. 

The Q-learning algorithm should generate Q-values for all state-action combinations if all combinations are explored training under a large number of times training. At each state, the action with the largest Q-value is chosen as the optimal. 

\subsection{Traffic Control with Q-learning}
The traffic light control with Q-learning intends to avoid congestion on each within the network by preventing the number of vehicles exceeding the capacity of each road. Assuming the initial number of vehicles on each road, the volume of influx vehicle and fluctuation of vehicle volume are given, the most critical issue is to determine the optimal green light passing time given the traffic flow states within the network. If the green light passing time is too short, congestion happens as the influx is larger than the exiting traffic volume; if the green light passing time is too long, the connecting road receive excessive traffic flow from the current road and the congestion is transferred to the connecting road. In the proposed scheme, the optimal green light passing time is determined using a Q-learning agent. The agent perceives the traffic state of the road network and selects the passing time at each intersect. The selected passing times in the network lead to the traffic state transition. The Q-learning agent calculates the reward for the state transition and the corresponding Q-value for the current state and action. The elements are as follows:
\subsubsection{State} It is hard to determine a state function that depicts the change of traffic flow under different control efforts. To increase calculation efficiency, the Q-learning agent uses discrete states without function approximator. The agent uses a state vector to represent the states of the traffic network, which are the numbers of the vehicle on each road. A state table is used to collect the state vectors and assess the traffic conditions in this control scenario. Because the learning time increases drastically with the size of the Q-table, we include only states that represent the traffic conditions on the main roads in the network. 
\begin{equation}
 S = \{s_{i}| i \in \mathbf{N}\}
\end{equation}
\begin{equation}
    s_{i} = [s_{i1},s_{i2}... s_{ij}],  s_{ij} \in \mathbf{N}
\end{equation}
$s_{i}$ refers to $i^{th}$ state, and $s_{ij}$ describes the number of vehicles (queue length) on the $j^{th}$ road in the $i^{th}$ time step.
\subsubsection{Action} The agent controls the traffic flow in the traffic network by adjusting the green light passing time at each intersection. Therefore, the green light passing time is the action by the agent. When implemented, the actions are in seconds of integer values in order to make discrete action space. Each action is represented with an action vector that includes green light passing times of all intersection. The size of the action space equals the multiple of numbers of passing time of all intersection. 
\begin{equation}
    A = \{a_{i}| i \in \mathbf{N}\}
\end{equation}
\begin{equation}
    a_{i} = [a_{i1},a_{i2}... a_{ij}], a_{ij} \in \mathbf{N}
\end{equation}
$a_{i}$ refers to $i^{th}$ action, and $a_{ij}$ controls the $j^{th}$ intersection in the $i^{th}$ action vector.

\subsubsection{reward} The objective of the Q-learning traffic control is to avoid congestion, aiming to minimize the vehicle queue lengths. When overflow happens, a punishment (negative value of the reward function) should be given to the agent. With that in mind, the reward function in this study can be define using a control cost function. We define the control cost function at time step k as a model:
\begin{equation}
    f_{k}(\mathbf{s},\mathbf{a}) =||\mathbf{s}_{k}||^{2} + \||\mathbf{a}_{k}-\mathbf{a}_{k-1}||^{2},
\end{equation}

where $\mathbf{s}_{k},\mathbf{a}_{k}$ denote the state vector, control vector, demand vector and disturbance vector at time step $k$ respectively. 

In order to reward the smooth ongoing of the traffic and punish the agent for making the queue length going over the capacity limit $\delta_{j}$ on the $j^{th}$ road, we define a punishment constant $P$ and an overflow vector $\mathbf{C} = [c_{1}.c_{2}...c_{j}]$, where 
\begin{equation}
    c_{kj}  = s_{kj} - \delta_{j}
\end{equation}

With that in mind, the reward function at time step k is determined as
\begin{equation}
    R_{k}(\mathbf{s},\mathbf{a}) = 
    \begin{cases}
    f_{k}(\mathbf{s},\mathbf{a}), c_{kj} \leq 0 \text{   for all $k,j \in \mathbf{N}$} \\
  
    -(f_{k}(\mathbf{s},\mathbf{a})+ R\cdot||\mathbf{C}_{k}||), \text{ if any $c_{kj} \geq 0$}

    \end{cases}
\end{equation}

Eq. (11) suggests that any control measures that lead to exceeding the road capacity will have a larger impact on the cumulative reward in Eq. (1) compared to the ones that successfully manages the network. The value of $R$ describes the extent to which the overflow traffic affects the value of the reward.

\section{Adaptive Horizon with Uncertainty}
\subsection{Traffic Control with Adaptive Horizon} Since the training time for all state-action pairs increases exponentially as the number of state-action pairs increases, we would like to determine the action space that yields the most efficient control outcome. In other words, the goal is to achieve the lowest control cost without examining a large action space, i.e.
\begin{equation}
    I = [min\{a_{ij}\},max\{a_{ij}\}]
\end{equation}
\begin{equation}
    \operatorname*{arg\,min}_{I} f(\mathbf{s},\mathbf{a})
\end{equation}
where $I$ denotes the interval of the $j^{th}$ intersection of the $i^{th}$ action (green light passing time).  
In order to approximate the global optimal solution, we start the training process by examining the action space with a larger range of green light passing time. From the result of the cost function, the interval of green light passing time in the action space that yields the minimum cost is selected as the new interval of the green light passing time, and the range of the actions is truncated. During this process, the dimension of the action space is kept constant while the resolution of the action space is increased, i.e. the number of values of light passing time for each intersection is constant and the difference between each value gradually decreases and the interval decreases. 
\begin{equation}
    [min\{a_{ij}\},max\{a_{ij}\}](m+1) \subseteq [min\{a_{ij}\},max\{a_{ij}\}](m)
\end{equation}
where $[min\{a_{ij}\},max\{a_{ij}\}](m)$ is equivalent to $I$ in Eq. (12) after the action space is trained $m$ times

As the set of actions converges to a certain interval, the agent start to experiment with the neighbouring values by shifting the green light passing time by a small amount, i.e. 
\begin{equation}
[min\{a_{ij}\},max\{a_{ij}\}](n+1) =[min\{a_{ij}\},max\{a_{ij}\}](n) \pm \eta
\end{equation}
assuming $m < n \in \mathbf{N}$, and $\eta \in \mathbf{N}$ is the leeway of adjusting the interval to obtain the local minimum cost function value.

\begin{figure}[h]
    \centering
    \includegraphics[width = 0.45\textwidth]{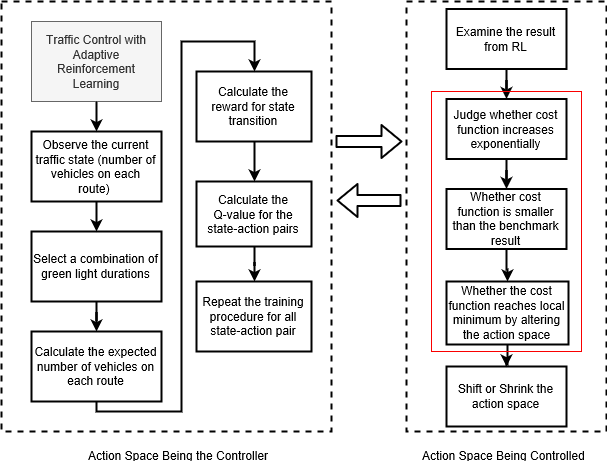}
    \caption{Program Flowchart}
    \label{fig:flowchart}
\end{figure}

The process of adapting the action space to the result of the cost function is iterative, as shown in Figure \ref{fig:flowchart}. The action space is at first a controlled variable when the general interval green light passing time is uncertain and becomes the controller of the traffic network once the action space converges to an interval that yields the minimum of the cost function.

\subsection{Traffic Control with Uncertainty}
The traffic control with uncertainty is presented in this section. Due to the various traffic patterns, uncertainty widely exits in traffic control problem. For instance, the traffic influx may vary due to the daily work hour. As we have come up with the traffic control strategy shown in previous sections, we wish to test whether the traffic controller is effective when traffic variance occurs. Assume that with historical data collected in the last few months and stored in the traffic database, the probability distribution function of traffic disturbance can be approximated with statistic tools, and the variation pattern can be determined. With the mean value $\mu_{1}$ and variance $\sigma_{1}$, the traffic demand $d_{j}$ of the $j_{th}$ road at time step $k$ can be defined with a normal distribution, 
\begin{equation}
    d_{i}(k) \sim N(\mu_{1},\sigma_{1}), k \in \mathbf{N}
\end{equation}

On the other hand, we define the turning rate to be $\tau_{j,p,i,r}$ to be the the proportion of vehicle that turns to the $r^{th}$ road of the $i^{th}$ intersection from the $p^{th}$ road of the $j^{th}$ intersection. The turning rate is also assumed to follow a normal distribution with  mean value $\mu_{2}$ and variance $\sigma_{2}$,
\begin{equation}
    \tau_{j,p,i,r}(k) \sim N(\mu_{2},\sigma_{2}), k \in \mathbf{N}
\end{equation}

We would like to increase the variance gradually and examine the robustness of the reinforcement learning algorithm under uncertain environment, by examining whether the agent can achieve low control cost, assuming that the states follow the equation
\begin{equation}
    s_{k+1} = s_{k}+Bu_{k} + Dd_{k}+e_k
\end{equation}
where $u_{k}$ is the control vector and $B$, $D$ are the control input matrix and demand matrix, respectively. We wish to identify the propagation of uncertainties to the control cost effort.

\section{Simulation Results}
\begin{figure}
    \centering
    \includegraphics[width = 0.4\textwidth]{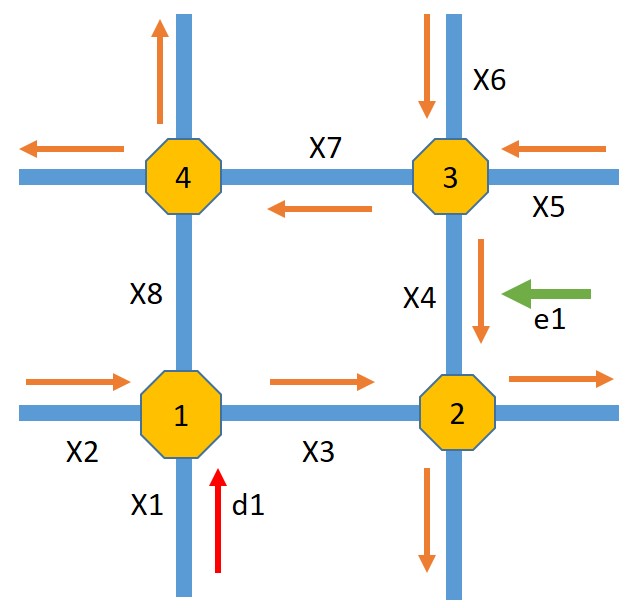}
    \caption{Simulated Network Diagram}
    \label{fig:network}
\end{figure}

In this paper, we perform simulation on the road network with four intersections connecting to 8 roads, as shown in Fig. \ref{fig:network}. The roads 1,2,5,6 are the paths from which traffic comes into the network, and others are the exiting paths for the traffic. The lengths of the input road are 600m, and the lengths of the rest are 700m. The basic parameters of the simulation are presented in Table \ref{tab:parameters} and the turning rates of the connecting roads within the network are presented in Table \ref{tab:turn}. All of the simulations are completed on a computer with 2.2 GHz Intel Core(TM) i7 Processor, with MATLAB R2015a software. 

Each intersection has only STOP and GO phases. Without loss of generality, we discretize the continuous dynamics into discrete time steps k, and we define the sampling time as T, the queue length on road r of the $i^{th}$ intersection within the $k^{th}$ time step as
\begin{equation}
    x_{i,r}(k+1) = x_{i,r}(k)+T\cdot(q^{in}_{i,r}(k)-q^{out}_{i,r}(k))+e_{i,r}(k) 
\end{equation}
where $q^{in}_{i,r}(k),q^{out}_{i,r}$ and $e_{i,r}(k)$ are the entering, exiting traffic rate and traffic disturbance (stochastic traffic fluctuation) of the $r^{th}$ road within the simulation interval $[(k_{i}-1)\cdot T_{i}, k_{i}\cdot T_{i}]$. 

The average entering and exiting traffic rate $q^{in}_{i,r}(k)$ and $q^{out}_{i,r}$ are updated based on the following equations
\begin{equation}
    q^{in}_{i,r}(k) =  \sum\tau_{j,p,i,r}(k)\cdot q^{out}_{j,p}(k)
\end{equation}
assuming that $p$ belongs to the set of incoming roads of road $r$, and
\begin{equation}
    q^{out}_{i,r}(k) = \frac{S_{i,r}}{C}\cdot u_{i,w}(k)
\end{equation}
where $S_{i,r}$ is the saturation of road $r$ at intersection $i$ and $C$ is the cycle time that equals the sum of passing and stopping time for vehicles on one direction. 

\begin{table}[]
    \centering
    \begin{tabular}{c   c}
        \hline
        \\Parameter & Value\\
        \hline
        Cycle time & 120s\\
        control interval length & 200s\\
        saturation flow rate & 3600 veh/h\\
        average vehicle length & 5m\\
        constraint on green time & 30 $\leq$ $a$ $\leq$ 90s\\
        $\epsilon$ & 0.5\\
        $\gamma$ & 0.9\\
    \end{tabular}
    \caption{Simulation Parameters}
    \label{tab:parameters}
\end{table}

\begin{table}[]
    \centering
    \begin{tabular}{c c c c c}  
         \hline
         \\ $\tau_{p,r}$ & x3 & x4 & x7 & x8\\
         \hline
         x1 & 0.3 & - & - & 0.7\\
         x2 & 0.8 & - & - & 0.2\\
         x5 & - & 0.6 & 0.4 & -\\
         x6 & - & 0.2 & - & 0.8\\
    \end{tabular}
    \caption{Network Turning Rate}
    \label{tab:turn}
\end{table}

In order to verify the effectiveness of the proposed reinforcement learning traffic control (abbreviated as Reg-RL) and the traffic control with reinforcement learning with the adaptive horizon (abbreviated as Adaptive-RL), the performance is compared with an MPC model solved by Sequential Quadratic Programming (abbreviated as MPC-SQP). \cite{ye2017stochastic} Both the solution and cost function values are presented and analyzed. In order to measure the performance of the algorithms, we present the average queue length of each road within the network of all time steps. To compare with the benchmark method, we present the control costs generated in the whole simulation. Last but not least, we investigate the fluctuation of the control costs under different levels of uncertainties and the impact of uncertainty on the system control costs. 

\begin{figure}[]
    \includegraphics[width = 0.45\textwidth]{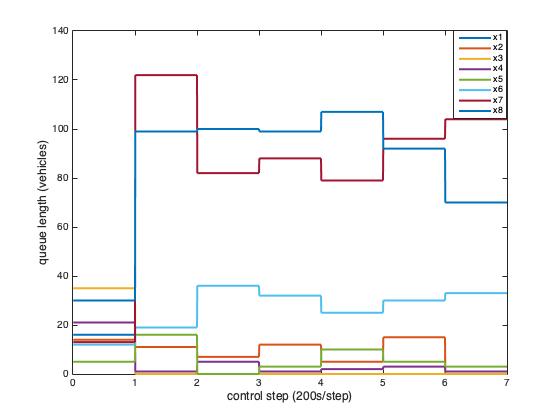}
    \caption{Queue Length with Regular RL}
    \label{fig:Reg_rl_queue}
\end{figure}

Figure \ref{fig:Reg_rl_queue} illustrates the queue length of all 8 links, from $X_{1}$ to $X_{8}$. Compared with the MPC-SQP method, the Reg-RL converges to a steady solution much quicker than the MPC-SQP method, and there is no traffic outflow (exceeding the upper bound) happening during the entire simulation period. This may be due to the positive reward that encourages the agent to select the action ones it determines the actions are effective and the emphasized punishment for exceeding the upper bound within reinforcement learning. Due to the sequence of the states during the training process, the agent tends to introduce the vehicle flow to the path for exiting the network ($X_{7}$ and $X_{8}$) while clearing up the body of the network. On the other hand, the result that the queue length is steady means the traffic condition is more predictable and avoids large variation in the traffic network. 

\begin{figure}[]
    \includegraphics[width = 0.45\textwidth]{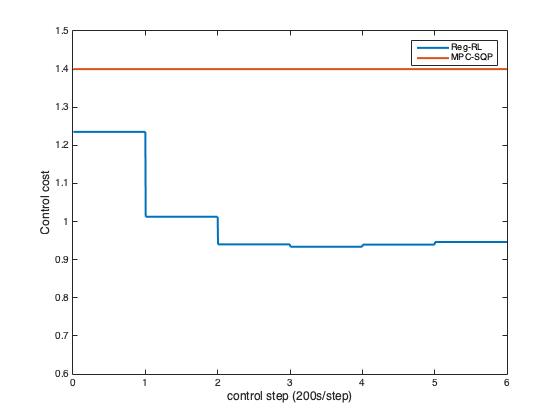}
    \caption{Cost of Regular RL}
    \label{fig:reg_rl_cost}
\end{figure}

From Figure \ref{fig:reg_rl_cost}, it can be seen that the value of the cost function gradually decreases throughout the simulation, with the simulation cost significantly lower than the value of the cost function of MPC-SQP. Similar to the queue length in Figure \ref{fig:Reg_rl_queue}, the value of cost function quickly converges to the steady state value during the simulation and significant fluctuation does not occur. This trend is quite different from the one in the MPC-SQP method, in which the control cost is negligible from the beginning but gradually increases as control steps increases. 

\begin{figure}[]
    \includegraphics[width = 0.45\textwidth]{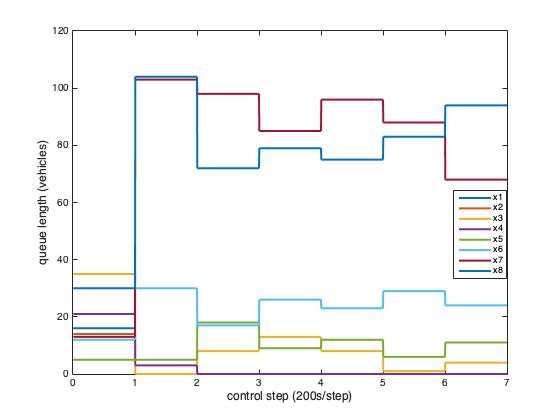}
    \caption{Queue Length with Adaptive Horizon}
    \label{fig:adaptivequeue}
\end{figure}

\begin{figure}[]
    \includegraphics[width = 0.45\textwidth]{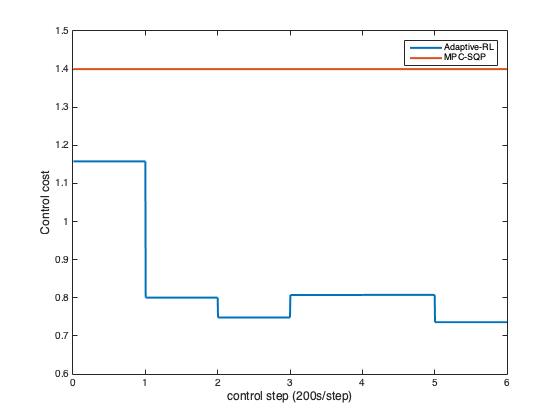}
    \caption{Cost of Adaptive Horizon}
    \label{fig:adaptivecost}
\end{figure}

The Adaptive-RL is based on the Reg-RL with the adaptive horizon method (varying action space). As seen in Figure \ref{fig:adaptivequeue}, the queue length at the network exit ($X_{7}$ and $X_{8}$) is smaller than the one in the Reg-RL method, meaning the exit section is further optimized compared to the one in the Reg-RL simulation. On the other hand, the value of the cost function of the Adaptive-RL method is also lower than the value of cost function in the Reg-RL method as shown in Figure \ref{fig:adaptivecost}. Both the queue length and the cost function value of the Adaptive-RL method follow the same trend as the queue length and the cost function value of the Reg-RL method. 

As training iteration increases, the span of action space decreases and the action for each intersection converges to an interval much smaller than initial constraint on green time presented in Table \ref{tab:parameters}. The  green passing time interval of intersection 1 is [55,65], [65,70] seconds for intersection 2, [65, 70] seconds for intersection 3 and [70, 75] seconds for intersection 4. The difference between green passing time values within time interval also decreases as the interval contracts, from 10 to 1. The final training time for 6 time-steps is approximately the same as the regular reinforcement learning. 

\begin{table}
\caption{Control Cost under Uncertain Influx}
\centering
\begin{tabular}{c c c c c c c}
\hline
Uncertainty & step 1 & step 2 & step 3 & step 4 & step 5 & step 6\\
\hline
5\% & 1.18 & 0.86 & 0.76 & 0.89 & 0.79 & 0.85\\
10\% & 1.02 & 0.89 & 0.95 & 0.93 & 0.87 & 0.83\\
20\% & 1.11 & 0.82 & 0.81 & 0.88 & 0.84 & 0.82\\
30\% & 1.11 & 0.80 & 0.80 & 0.79 & 0.74 & 0.84\\
40\% & 1.27 & 0.99 & 149.19 & 279.88 & 149.92 & 10.25\\
\end{tabular}
\label{tab}
\end{table}

To test the robustness of the RL controller with the adaptive horizon, we introduced uncertainty in the influx of traffic to the simulation with increasing variance. The $\%$ Uncertainty in Table \ref{tab} is the percentage of variance relative to the influx volume. As shown in Table \ref{tab}, when uncertainty is lower than 30$\%$, the agent is able to achieve low control cost. As uncertainty increases to 40$\%$, control cost increases dramatically, which means the traffic network is overcrowded and the controller fails to control the network. From here it can be said that the room for tolerating the traffic influx variance is decent. 

\begin{table}[t]
\caption{Control Cost under Uncertainty Turning rate}
\centering
\begin{tabular}{c c c c c c c}
\hline
Uncertainty & step 1 & step 2 & step 3 & step 4 & step 5 & step 6\\
\hline
5\% & 0.79 & 0.46 & 0.39 & 0.50 & 0.52 & 0.66\\
10\% & 0.62 & 0.28 & 0.31 & 0.38 & 0.42 & 0.47\\
15\% & 0.69 & 0.38 & 0.35 & 0.37 & 0.31 & 0.34\\
20\% & 0.85 & 0.62 & 0.84 & 0.92 & 30.55 & 0.60\\
30\% & 408.4 & 57.69 & 1.51 & 94.97 & 0.75 & 1.53\\
\end{tabular}
\label{tab:turnvar}
\end{table}

On the other hand, we also want to investigate the robustness of the RL controller with adaptive horizon under uncertain turning rates. Same as in Table \ref{tab}, the $\%$ Uncertainty in Table \ref{tab:turnvar}  is the variance relative to the turning rates in Table \ref{tab:turn}. When uncertainty is under 15\%, the agent is able to control the network with a control cost significantly lower than the MPC-SQP method. When uncertainty goes above 20\%, control cost increases exponentially, which means that traffic overflow occurs. A closer investigation suggests that road 3 is mostly affected and easily leads to traffic overflow. The tolerance of turning rate to uncertainty is lower than the one of traffic disturbance.

\section{Conclusion}
This paper investigates a reinforcement learning approach for traffic control with the adaptive horizon. In order to reduce the computation time and improve the efficiency of the controller, the action space of the Q-learning algorithm gradually converges towards a subset of the original action space that is most likely to obtain optimal control outcome. The Q-learning based controller is able to reach a stable solution in a shorter time with lower control cost. Meanwhile, the controller is robust under uncertain traffic demand and turning rate.

In the future, testing with additional sources of uncertainty should be investigated. Deep Q-learning can also be combined with the adaptive horizon to expand the research to large state space. Furthermore, the convergence of the Q-learning algorithm under uncertainty can also be analyzed. 

\section*{Acknowledgments}
We gratefully acknowledge the support from the National Science Foundation (DMS-1555072, DMS-1736364, and DMS-1821233).

\appendices

\ifCLASSOPTIONcaptionsoff
  \newpage
\fi



%


\bibliographystyle{IEEEtran}
\bibliography{ref}

\begin{thebibliography}{10}
\providecommand{\url}[1]{#1}
\csname url@samestyle\endcsname
\providecommand{\newblock}{\relax}
\providecommand{\bibinfo}[2]{#2}
\providecommand{\BIBentrySTDinterwordspacing}{\spaceskip=0pt\relax}
\providecommand{\BIBentryALTinterwordstretchfactor}{4}
\providecommand{\BIBentryALTinterwordspacing}{\spaceskip=\fontdimen2\font plus
\BIBentryALTinterwordstretchfactor\fontdimen3\font minus
  \fontdimen4\font\relax}
\providecommand{\BIBforeignlanguage}[2]{{%
\expandafter\ifx\csname l@#1\endcsname\relax
\typeout{** WARNING: IEEEtran.bst: No hyphenation pattern has been}%
\typeout{** loaded for the language `#1'. Using the pattern for}%
\typeout{** the default language instead.}%
\else
\language=\csname l@#1\endcsname
\fi
#2}}
\providecommand{\BIBdecl}{\relax}
\BIBdecl

\bibitem{gartner1995development}
N.~H. Gartner, C.~Stamatiadis, and P.~J. Tarnoff, ``Development of advanced
  traffic signal control strategies for intelligent transportation systems:
  Multilevel design,'' \emph{Transportation Research Record}, no. 1494, 1995.

\bibitem{singh2009time}
L.~Singh, S.~Tripathi, and H.~Arora, ``Time optimization for traffic signal
  control using genetic algorithm,'' \emph{International Journal of Recent
  Trends in Engineering}, vol.~2, no.~2, p.~4, 2009.

\bibitem{yang2016isolated}
K.~Yang, S.~I. Guler, and M.~Menendez, ``Isolated intersection control for
  various levels of vehicle technology: Conventional, connected, and automated
  vehicles,'' \emph{Transportation Research Part C: Emerging Technologies},
  vol.~72, pp. 109--129, 2016.

\bibitem{papageorgiou2003review}
M.~Papageorgiou, C.~Diakaki, V.~Dinopoulou, A.~Kotsialos, and Y.~Wang, ``Review
  of road traffic control strategies,'' \emph{Proceedings of the IEEE},
  vol.~91, no.~12, pp. 2043--2067, 2003.

\bibitem{ye2016hierarchical}
B.-L. Ye, W.~Wu, L.~Li, and W.~Mao, ``A hierarchical model predictive control
  approach for signal splits optimization in large-scale urban road networks,''
  \emph{IEEE Transactions on Intelligent Transportation Systems}, vol.~17,
  no.~8, pp. 2182--2192, 2016.

\bibitem{chen2016improved}
S.~Chen and D.~J. Sun, ``An improved adaptive signal control method for
  isolated signalized intersection based on dynamic programming,'' \emph{IEEE
  Intelligent Transportation Systems Magazine}, vol.~8, no.~4, pp. 4--14, 2016.

\bibitem{xie2012schedule}
X.-F. Xie, S.~F. Smith, L.~Lu, and G.~J. Barlow, ``Schedule-driven intersection
  control,'' \emph{Transportation Research Part C: Emerging Technologies},
  vol.~24, pp. 168--189, 2012.

\bibitem{kong2011urban}
X.~Kong, G.~Shen, F.~Xia, and C.~Lin, ``Urban arterial traffic two-direction
  green wave intelligent coordination control technique and its application,''
  \emph{International Journal of Control, Automation and Systems}, vol.~9,
  no.~1, pp. 60--68, 2011.

\bibitem{lu2011two}
K.~Lu, X.~Zeng, L.~Li, and J.~Xu, ``Two-way bandwidth maximization model with
  proration impact factor for unbalanced bandwidth demands,'' \emph{Journal of
  Transportation Engineering}, vol. 138, no.~5, pp. 527--534, 2011.

\bibitem{wang2010parallel}
F.-Y. Wang, ``Parallel control and management for intelligent transportation
  systems: Concepts, architectures, and applications,'' \emph{IEEE Transactions
  on Intelligent Transportation Systems}, vol.~11, no.~3, 2010.

\bibitem{aboudolas2010rolling}
K.~Aboudolas, M.~Papageorgiou, A.~Kouvelas, and E.~Kosmatopoulos, ``A
  rolling-horizon quadratic-programming approach to the signal control problem
  in large-scale congested urban road networks,'' \emph{Transportation Research
  Part C: Emerging Technologies}, vol.~18, no.~5, pp. 680--694, 2010.

\bibitem{kaelbling1996reinforcement}
L.~P. Kaelbling, M.~L. Littman, and A.~W. Moore, ``Reinforcement learning: A
  survey,'' \emph{Journal of artificial intelligence research}, vol.~4, pp.
  237--285, 1996.

\bibitem{el2014design}
S.~El-Tantawy, B.~Abdulhai, and H.~Abdelgawad, ``Design of reinforcement
  learning parameters for seamless application of adaptive traffic signal
  control,'' \emph{Journal of Intelligent Transportation Systems}, vol.~18,
  no.~3, pp. 227--245, 2014.

\bibitem{li2017reinforcement}
Z.~Li, P.~Liu, C.~Xu, H.~Duan, and W.~Wang, ``Reinforcement learning-based
  variable speed limit control strategy to reduce traffic congestion at freeway
  recurrent bottlenecks,'' \emph{IEEE transactions on intelligent
  transportation systems}, vol.~18, no.~11, pp. 3204--3217, 2017.

\bibitem{aslani2018traffic}
M.~Aslani, S.~Seipel, M.~S. Mesgari, and M.~Wiering, ``Traffic signal
  optimization through discrete and continuous reinforcement learning with
  robustness analysis in downtown tehran,'' \emph{Advanced Engineering
  Informatics}, vol.~38, pp. 639--655, 2018.

\bibitem{mannion2015parallel}
P.~Mannion, J.~Duggan, and E.~Howley, ``Parallel reinforcement learning for
  traffic signal control,'' \emph{Procedia Computer Science}, vol.~52, pp.
  956--961, 2015.

\bibitem{wei2018intellilight}
H.~Wei, G.~Zheng, H.~Yao, and Z.~Li, ``Intellilight: A reinforcement learning
  approach for intelligent traffic light control,'' in \emph{Proceedings of the
  24th ACM SIGKDD International Conference on Knowledge Discovery \& Data
  Mining}.\hskip 1em plus 0.5em minus 0.4em\relax ACM, 2018, pp. 2496--2505.

\bibitem{huang2018deep}
X.~Huang, T.~Yuan, G.~Qiao, and Y.~Ren, ``Deep reinforcement learning for
  multimedia traffic control in software defined networking,'' \emph{IEEE
  Network}, vol.~32, no.~6, pp. 35--41, 2018.

\bibitem{watkins1992q}
C.~J. Watkins and P.~Dayan, ``Q-learning,'' \emph{Machine learning}, vol.~8,
  no. 3-4, pp. 279--292, 1992.

\bibitem{ye2017stochastic}
B.-L. Ye, W.~Wu, H.~Gao, Y.~Lu, Q.~Cao, and L.~Zhu, ``Stochastic model
  predictive control for urban traffic networks,'' \emph{Applied Sciences},
  vol.~7, no.~6, p. 588, 2017.

\end{thebibliography}
%

\end{document}